\documentclass[amsmath,amssymb,twocolumn]{revtex4-1}
\usepackage{graphicx}
\usepackage{hyperref}
\usepackage{url}
\usepackage{subfigure}
\begin{document}

\title{The end-to-end testbed of the Optical Metrology System on-board LISA Pathfinder}
\author{Frank Steier} \email[Corresponding author:]{frank.steier@aei.mpg.de}
\author{Felipe Guzm\'an Cervantes} \email[Corresponding author:]{felipe.guzman@aei.mpg.de}
\author{Antonio F. Garc\'ia~Mar\'in}
\author{Gerhard Heinzel}
\author{Karsten Danzmann}
\affiliation{Max-Planck-Institut f\"ur Gravitationsphysik (Albert-Einstein-Institut) and Universit\"at Hannover, Germany}
\author{Domenicio Gerardi}
\affiliation{EADS Astrium Satellites GmbH, Friedrichshafen, Germany}
\begin{abstract}
LISA Pathfinder is a technology demonstration mission for the Laser Interferometer Space Antenna (LISA). The main experiment on-board LISA Pathfinder is the so-called LISA Technology Package (LTP) which has the aim to measure the differential acceleration between two free-falling test masses with an accuracy of $3\times10^{-14}$\,ms$^{-2}$/$\sqrt{\rm Hz}$ between 1\,mHz and 30\,mHz. This measurement is performed interferometrically by the Optical Metrology System (OMS) on-board LISA Pathfinder. In this paper we present the development of an experimental end-to-end testbed of the entire OMS. It includes the interferometer and its sub-units, the interferometer back-end which is a phasemeter and the processing of the phasemeter output data. Furthermore, 3-axes piezo actuated mirrors are used instead of the free-falling test masses for the characterisation of the dynamic behaviour of the system and some parts of the Drag-free and Attitude Control System (DFACS) which controls the test masses and the satellite.
The end-to-end testbed includes all parts of the LTP that can reasonably be tested on earth without free-falling test masses. At its present status it consists mainly of breadboard components. Some of those have already been replaced by Engineering Models of the LTP experiment. In the next steps, further Engineering Models and Flight Models will also be inserted in this testbed and tested against well characterised breadboard components. The presented testbed is an important reference for the unit tests and can also be used for validation of the on-board experiment during the mission.
\end{abstract}

\maketitle
\section{Introduction}

LISA Pathfinder is presently in its implementation phase \cite{paul}. Its core experiment, LTP, has passed the Critical Design Review which is a major milestone in the development phase with the consequence that the design of all sub-units is frozen. The sensitivity of the interferometer including its sub-units was already demonstrated on breadboard level with a straight-forward data processing of its output data \cite{ltpperf}. On-board LISA Pathfinder this processing includes various additional features that are necessary in order to increase the reliability of the experiment or for technical reasons due to the system design. Those features are:
\begin{itemize}
\item the averaging of nominal and redundant interferometer outputs,
\item the asynchronous data transfer of the science data, including downsampling, between the OMS back-end, called the Data Management Unit (DMU), and the On-board Computer (OBC),
\item the task separation between the DMU and the OBC for the initial test mass alignment on orbit and
\item the implementation of a suitable real-time error detection and propagation scheme that is performed on the DMU input data.
\end{itemize}

Apart from the error detection, each of these features is implemented in the end-to-end testbed which was used to demonstrate the performance of the entire OMS. The testbed contains breadboard components of the LTP units and the Engineering Model of the optical bench. The test masses were substituted by piezo actuated dummy mirrors that were used to characterise the response of the interferometer to dynamical test masses and to measure the residual cross-talk of different OMS output channels. These piezos have sufficient stability to reach the required interferometer sensitivity \cite{feli}. Thus, the testbed includes all components of the OMS on-board LISA Pathfinder that can reasonably be tested on ground if a system with full performance is needed.

In the near future other Engineering and Flight Models will be available and can be tested within this setup and the results presented here can be used as reference for those tests. With both breadboard components or Engineering or spare Flight Models the testbed can later be used as end-to-end simulator for the experiment during the LISA Pathfinder mission. One possible enhancement for the experiment in the future is to include DFACS laws and a dynamical model of LTP in the piezo control.

\section{Experimental Setup}

\begin{figure*}[htb]
\centering
\includegraphics[width=0.85\textwidth]{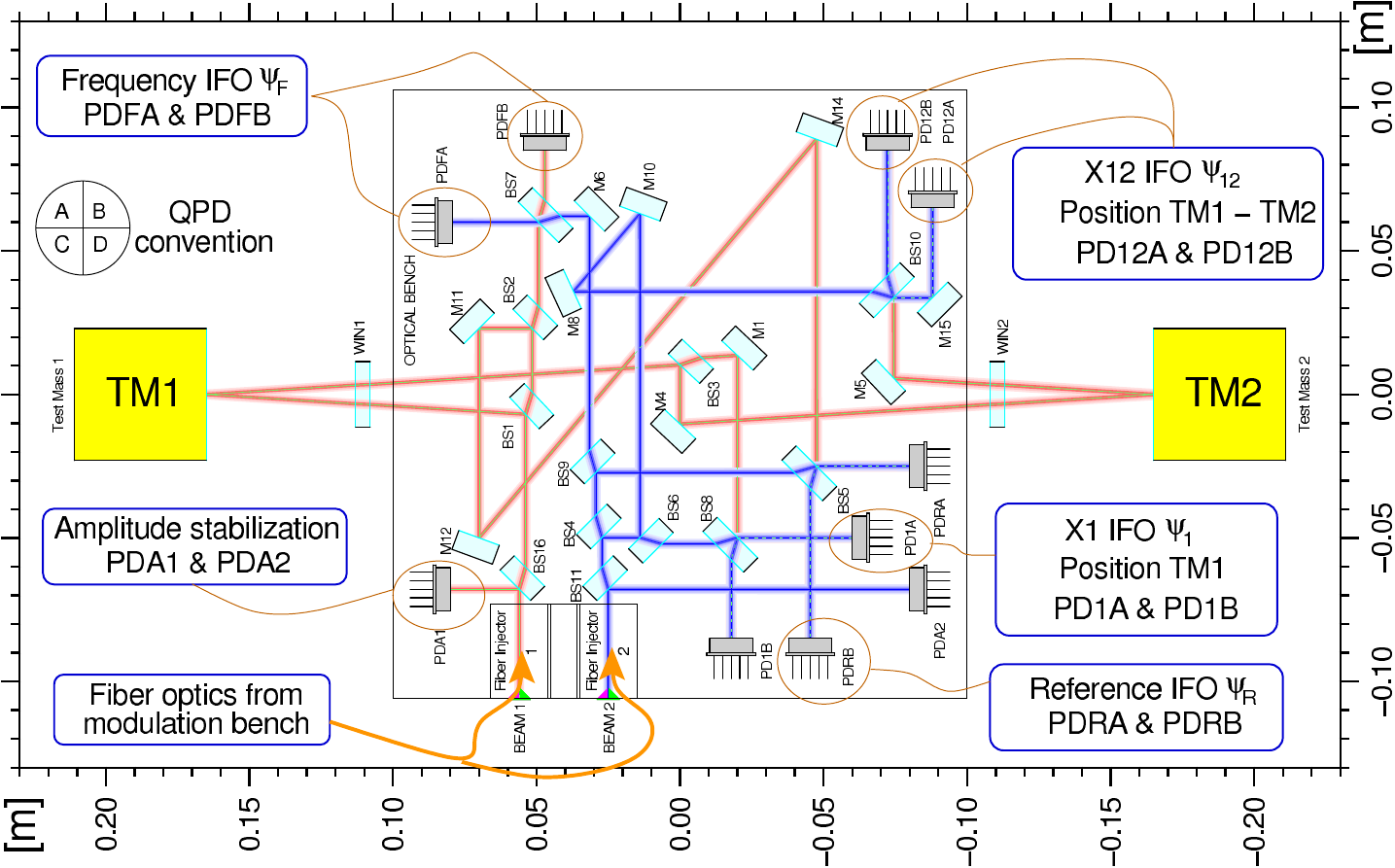}
\caption{Schematic layout of the Engineering Model of the LTP interferometer.}
\label{OB}
\end{figure*}

The core part of the optical system on-board LISA Pathfinder is the interferometer and the phasemeter. A schematic view of the interferometer can be seen in Figure \ref{OB}. It is the Engineering Model of the optical bench which is presently being used in the end-to-end testbed. It includes four interferometers. Two of them are used to sense the test mass positions: The so-called X1 interferometer measures the position of test\,mass\,1 with respect to the optical bench, and the X12 interferometer measures the differential displacement of both test masses \cite{laserfreq}.

\begin{figure*}[htb]
\centering
\includegraphics[width=0.95\textwidth]{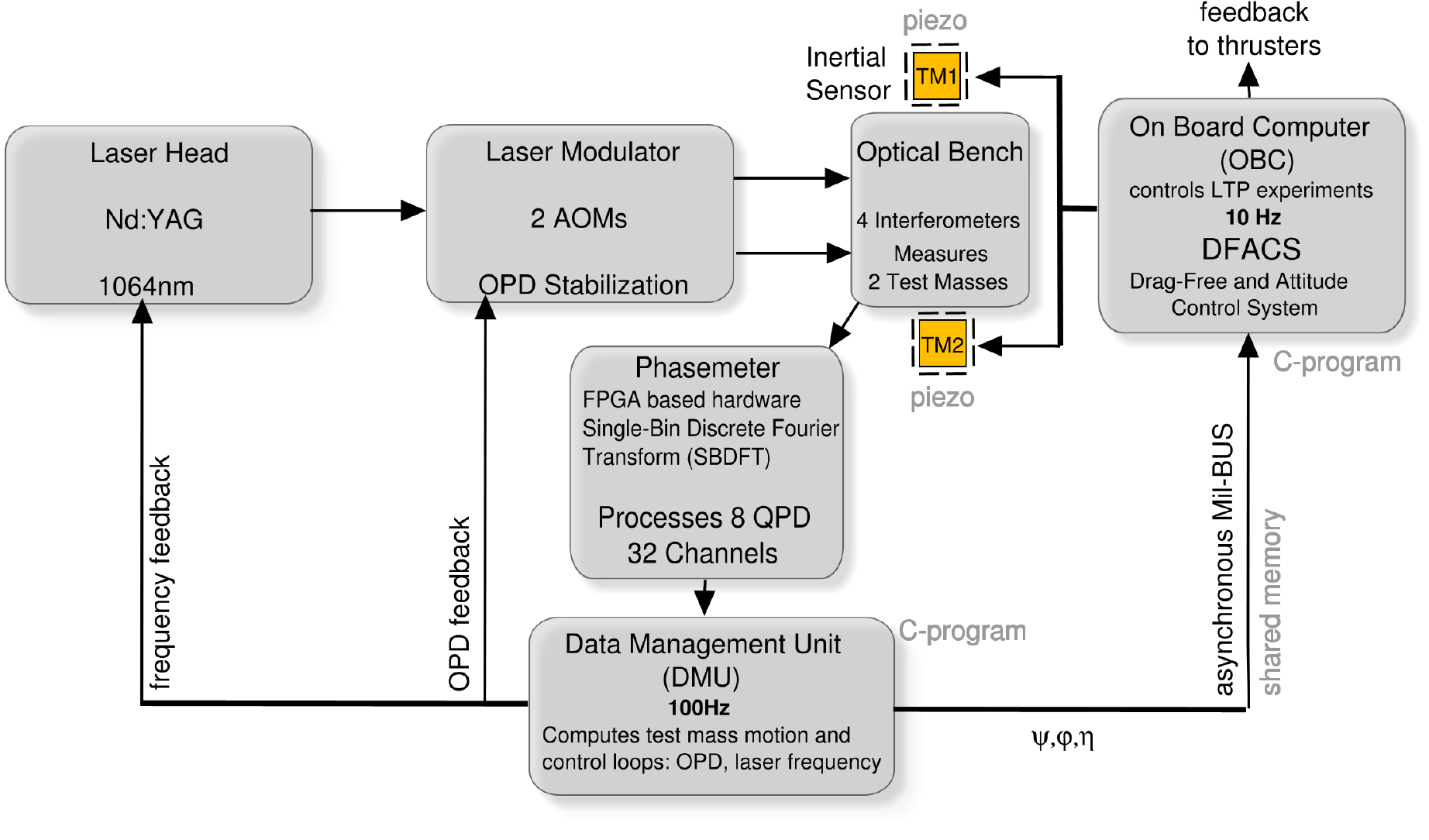}
\caption{Simplified view of the Drag-Free and Attitude Control System loop.}
\label{dfacs}
\end{figure*}

The LTP interferometer uses a heterodyne modulation and readout scheme. Two frequency shifted beams are generated in the Laser Modulator from an input laser beam and injected onto the optical bench where they are combined at the recombination beamsplitter of each interferometer. The beat signal is at the heterodyne frequency of about 1\,kHz and detected by two redundant quadrant photodiodes (QPDs) at the two output ports of this beamsplitter \cite{noisesources}. The photo currents produced are converted into a voltage and digitised for processing inside the phasemeter which is shown in Figure~\ref{dfacs} within the DFACS. The phasemeter calculates a DC voltage, $DC$, and a complex vector, $F$, at the heterodyne frequency via Single Bin Discrete Fourier Transform for each quadrant of each photodiode \cite{laserfreq}. The data is sent to the Data Management Unit and, there, combined to a single longitudinal measurement, $\Psi$, and two angular measurements, $\varphi$ and $\eta$, for each test mass. The angular readout is performed by calculating the beam centre on the QPDs using the DC signals and, additionally, by calculating the phase difference at the heterodyne frequency on the different halves of the QPDs. The latter measurement is called Differential Wavefront Sensing (DWS) and is highly sensitive around a zero phase difference, whereas the DC readout has a wider range but less sensitivity. These calculations are performed at a frequency of 100\,Hz and represent the main science output data of the experiment. The data is downsampled to 10\,Hz and sent to the On-Board Computer which has an asynchronous data link to the DMU. The OBC software includes the DFACS algorithm which controls the test masses via a combined capacitive sensor and actuator, called the Inertial Sensor, and the satellite via Micro-Newton thrusters.

The flow diagram in Figure \ref{dfacs} shows the LTP experiment. The gray labels indicate how the functionalities are implemented in the end-to-end testbed like for the 3-axes piezo actuated mirrors that are used instead of the free-falling test masses.
\subsection{Averaging of redundant readout channels}
The phasemeter output data transferred to the DMU includes two redundant sets of quadrant photodiode channels. In order to increase the signal-to-noise ratio of the science output data and to improve the reliability of the system in case of temporary or permanent failures in the readout channels, the redundant information is averaged. In principle two scenarios are possible for the averaging of the redundant data: One is to calculate the output for each QPD individually and to average the resulting test mass positions and angles. The second option is to average the information of the individual quadrants of the photodiodes and then use the results for further processing as if they were derived from a single quadrant photodiode.

\begin{figure}[htb]
\centering
\includegraphics[width=0.8\columnwidth]{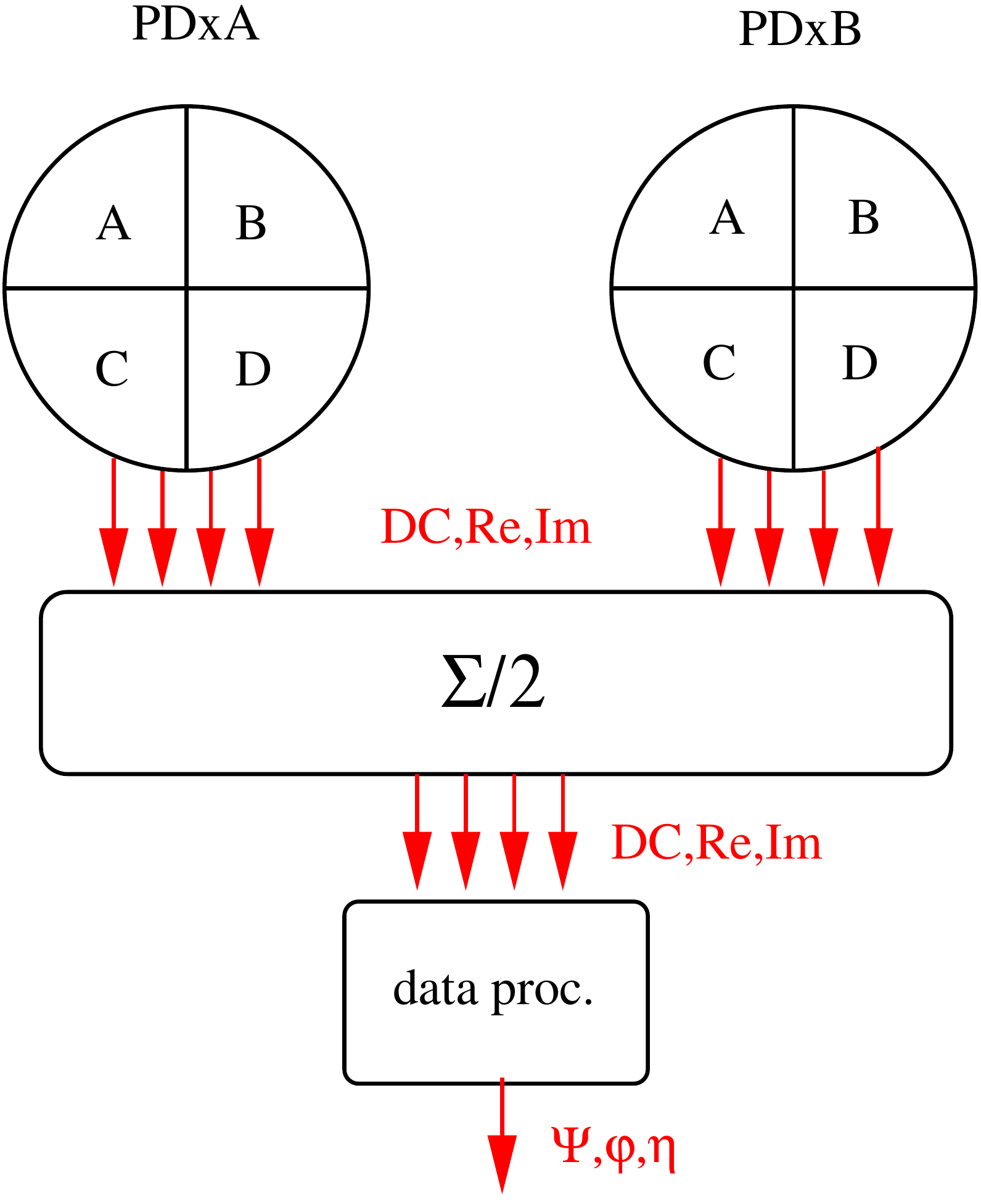}
\caption{Averaging of two redundant quadrant photodiodes.}
\label{average}
\end{figure}

The second option, illustrated in Figure \ref{average}, was chosen to be implemented. The DC output and the two components of the complex vector, \lq Re' and \lq Im', are averaged for each individual quadrant before further processing is performed. This has the advantage that more combinations of channel failures are acceptable: If, for example, the left half of one QPD, PDxA, and the right half of the redundant QPD,  PDxB, are erroneous, there are still four independent quadrants available that lead to a reasonable angular calculation. If the first option of the averaging was implemented, none of the individual angular calculations for PDxA and PDxB would lead to a meaningful result. An additional advantage of the second option is that specific straylight effects cancel in the longitudinal output data \cite{roland,meinediss}.

The actual implementation of the averaging has one important disadvantage. The differential alignment of two redundant photodiodes has an influence on the data processing. If, for example, both redundant QPDs have a different angle of rotation around the incoming beam, the nominally redundant quadrants sense slightly different parts of the interference pattern, which has an influence on the angular calculations. Using the first option, this effect could be compensated. In the actual implementation, however, a more accurate alignment of the photodiodes is necessary.

The averaged information was not used in first breadboard experiments, but is foreseen on-board LISA Pathfinder and implemented in the end-to-end testbed.

\subsection{Asynchronous data link}
As indicated in Figure \ref{dfacs}, the data link between DMU and OBC is asynchronous, and the DMU science output data must be downsampled before being transferred to the OBC. This has two major influences: One is the impact on the interferometer sensitivity and the other is the delayed delivery of the science output data which is used by the DFACS for the test mass control as described below. The aim of the investigations presented here was to minimise the delay by choosing an adequate anti-aliasing filter and to demonstrate the system performance with those filters and an asynchronous data link.

The DMU internal processing is performed at 100\,Hz and the output science data is transferred to the OBC at 10\,Hz. This requires downsampling by a factor of 10. The anti-aliasing lowpass filters that are needed for the downsampling must be efficient in terms of processing time due to the limitted capacity of the DMU. Two digital filters have been found for this purpose: an Infinite Impulse Response (IIR) filter and a Finite Impulse response (FIR) filter. The former has 6 coefficients which are multiplied by the recent input and output data of the filter at the input data rate of 100\,Hz. The latter is a simple average of the last 10 data samples that can be performed at the output data rate of 10\,Hz.

In order to achieve sufficiently large control bandwidth for the test mass control, the DFACS input data delay must be kept small. This delay is influenced on one hand by the processing time of the DMU and the phasemeter and on the other hand by the asynchronous data link. The former can be minimised by a suitable anti-aliasing filter design. The latter is minimised by a pseudo-synchronisation of the DMU and the OBC. A detailed description of the data link implementation on-board LISA Pathfinder can be found in \cite{meinediss}. The filter baseline design for the flight software is the moving average filter. It will finally be confirmed during the system tests with the representative Engineering and Flight Models.
 
On-board LISA Pathfinder the connection between DMU and OBC is implemented as a Mil-BUS. It is a modified serial bus developed for military application and nowadays also used in the satellite on-board subsystem data handling. In the actual real-time testbed the operations of the DMU and the OBC are taken over by two different C-programs and the simulation of the asynchronous data link is realised by a shared memory that is asynchronously written and read by the two C-programs.

\subsection{Optical Metrology System error handling}

The error handling of the Optical Metrology System is an essential new element in the data processing compared to earlier breadboard experiments. It consists of two parts: One part consits of the adapted calculation rules for the science data and the second part is the real-time classification of the DMU science output data quality. We defined an error handling and propagation scheme and investigated the influence of channel failures in the end-to-end testbed. The data quality classification will be implemented in the software of the Engineering Model DMU and tested within the presented testbed. However, in the present implementation of the testbed the data quality classification is not considered, since it includes many aspects that are specific to the Mil-BUS. 

The averaging of two redundant readout channels described above must be modified if failures occur in some of the data channels. If one channel of one redundant quadrant pair shows an anomaly, the other available channel still delivers the same information. Therefore, it is weighted twice, and the further processing is performed as usual. In case of more channel failures, the calculation rules of the science data must slightly be changed. A detailed description of the adapted calculation rules needed for this purpose can be found in \cite{meinediss}.

The real-time classification of the science output data is done in two steps. In every processing step 13 error checks are performed on each of the 32 phasemeter output channels. These checks can individually be disabled depending on the operational mode of the experiment. A channel selection logic is foreseen that decides which channels to use for the further processing. Its output and the interferometer contrasts are used to determine the data quality of the science output data which can have four different states. The quality classification logic is quite complex and, therefore, not described in detail here, but further information can be found in \cite{meinediss}.

\section{Experimental results}
Within the end-to-end testbed, important specific aspects of the data processing have been investigated. These are the demonstration of the performance with asynchronous data transfer, averaging of redundant photodiode channels and downsampling. Additionally, it was possible to investigate the residual interferometer readout cross-talk of the angular measurements and to test the on-orbit test mass alignment using the 3-axes piezo actuated mirrors.

\subsection{System performance}

\begin{figure}[htb]
\centering
\includegraphics[width=\columnwidth]{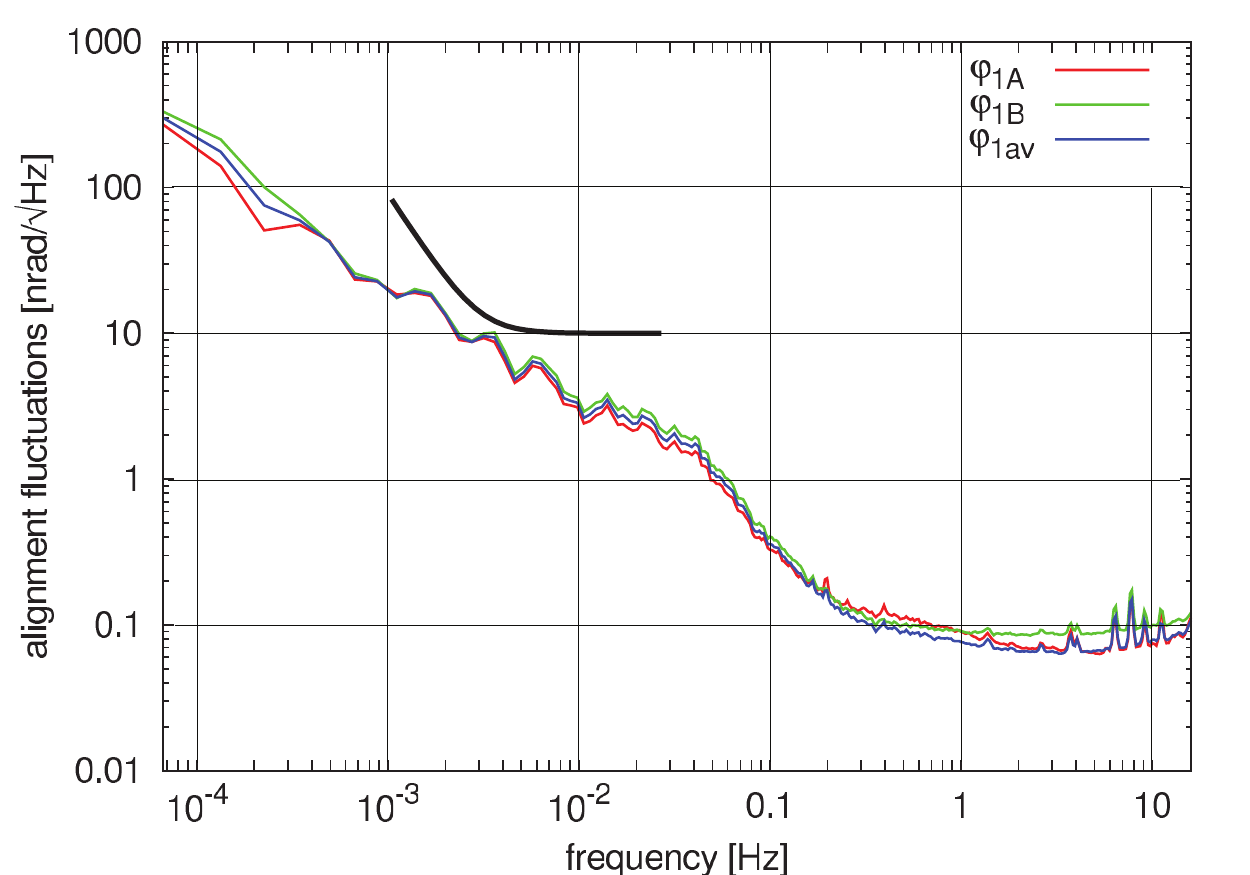}
\caption{Performance of the angular measurement for one angle of test\,mass\,1.}
\label{angperf}
\end{figure}

Two aspects of the data processing might have influence on the system performance: the averaging and the downsampling and asynchronous data transfer. For the longitudinal and angular readout of both test masses, the change in the sensitivity is similar if these features are included in the processing. As an example, the angular readout noise of one angle of test\,mass\,1 is shown in Figure \ref{angperf}. The individual calculations, $\varphi_{\rm 1A}$ and $\varphi_{\rm 1B}$, and the averaged result, $\varphi_{\rm 1av}$, are almost on the same level and well below the required noise budget of 10\,nrad/$\sqrt{\rm Hz}$.

\begin{figure}[htb]
\centering
\includegraphics[angle=-90,width=\columnwidth]{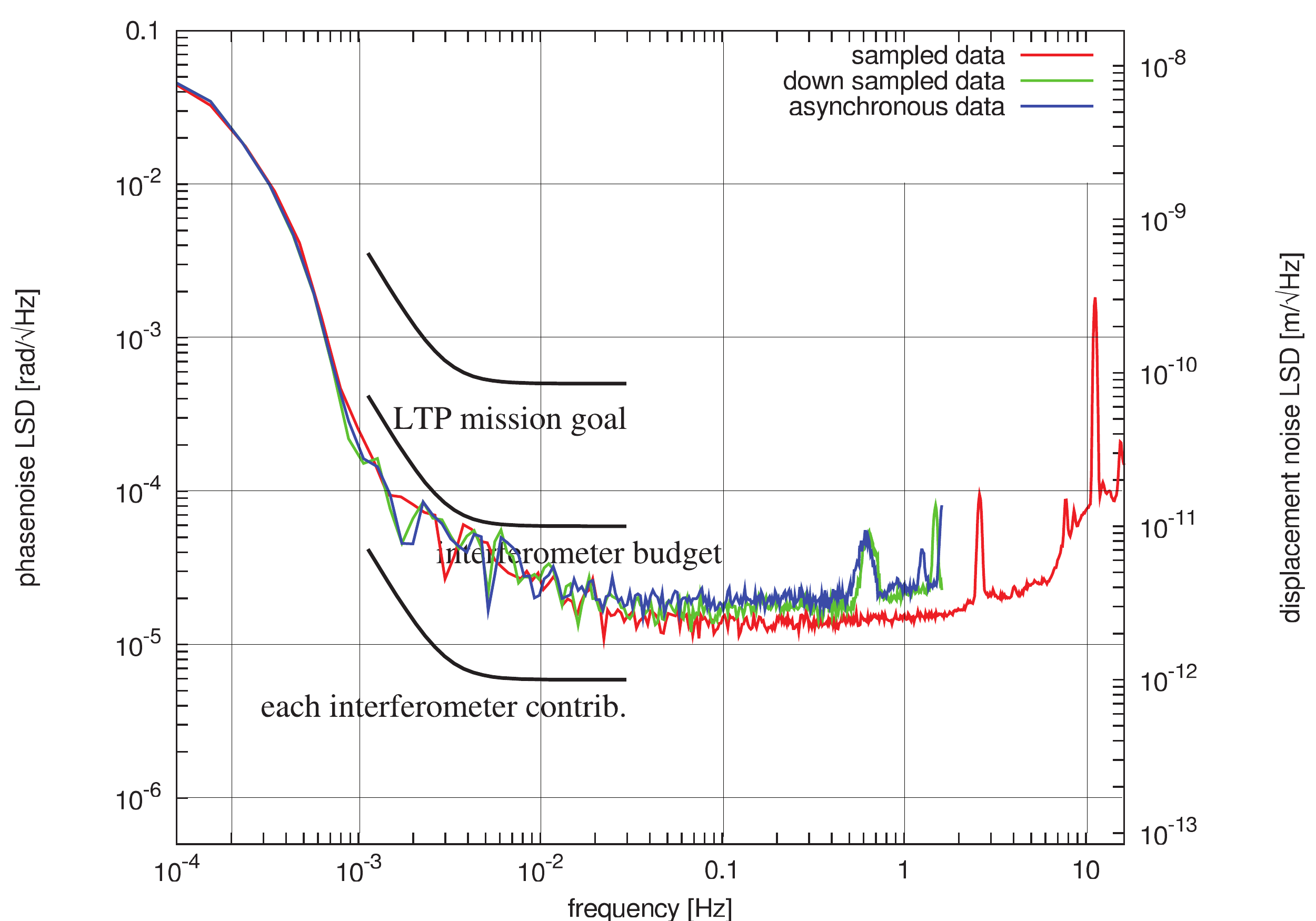}
\caption{Performance of the longitudinal measurement in the X1 interferometer.}
\label{longperf}
\end{figure}

The same was done for the longitudinal measurements. The sensitivity achieved with averaged signals is shown as red curve in Figure \ref{longperf} for the X1 interferometer. In addition, the synchronously downsampled data (green curve) and the asynchronously downsampled data (blue curve) are shown in this figure. The filter used here was an IIR filter with 6 coefficients. The high sampling rate was 32\,Hz in the experiment instead of 100\,Hz on-board LISA Pathfinder. However, the data was also reduced by a factor of 10 as in LTP. Down to approximately 100\,mHz the effect of aliasing is visible. However, in the measurement band between 1\,mHz and 30\,mHz it is negligible. At the higher sampling rate used on-board LISA Pathfinder, aliasing will appear at even higher frequencies. Deviations between the synchronous and asynchronous data can be seen at the characterstics peaks in the spectra. They are caused by the asynchronism which slightly shifts the spectra with respect to each other. The sensitivity of the system is well below the required interferometer goal of 10\,pm/$\sqrt{\rm Hz}$.

The X12 interferometer sensitivity is not shown in Figure \ref{longperf}, but would be on the same sensitivity level as the X1 interferometer, since we do not use real test masses but only fixed dummy mirrors. However, at the time the presented measurements were performed, the X12 interferometer was used with unstable piezo actuated mirrors and, therefore, the sensitivity reached was not representative.

In addition to the real-time simulation described above, a post-processing analysis that takes realistic clock deviations of the DMU and the OBC into account was performed \cite{meinediss}. In these simulations the 10 sample moving average filters were used. We showed that the use of these filters has also a negligible impact on the system performance.

\subsection{Residual angular cross-talk}
One of the most important experiments that are performed during the LISA Pathfinder mission is the measurement and compensation of residual cross-talk between different Degrees of Freedom of the test masses and the spacecraft. The interferometer itself may have cross-talk between its different angular readout channels. It was possible to set an upper limit of this cross-talk using the 3-axes piezo actuated mirrors. An orthogonalisation scheme was implemented in the hardware of the piezo drivers and, additionally, in software by linear combination of the processed angles. This way the piezo axes were actuated such that the commanded angular variations referred to the angles measured by the interferometer.

\begin{figure*}[!htb]
\centering
\includegraphics[angle=-90,width=0.89\textwidth]{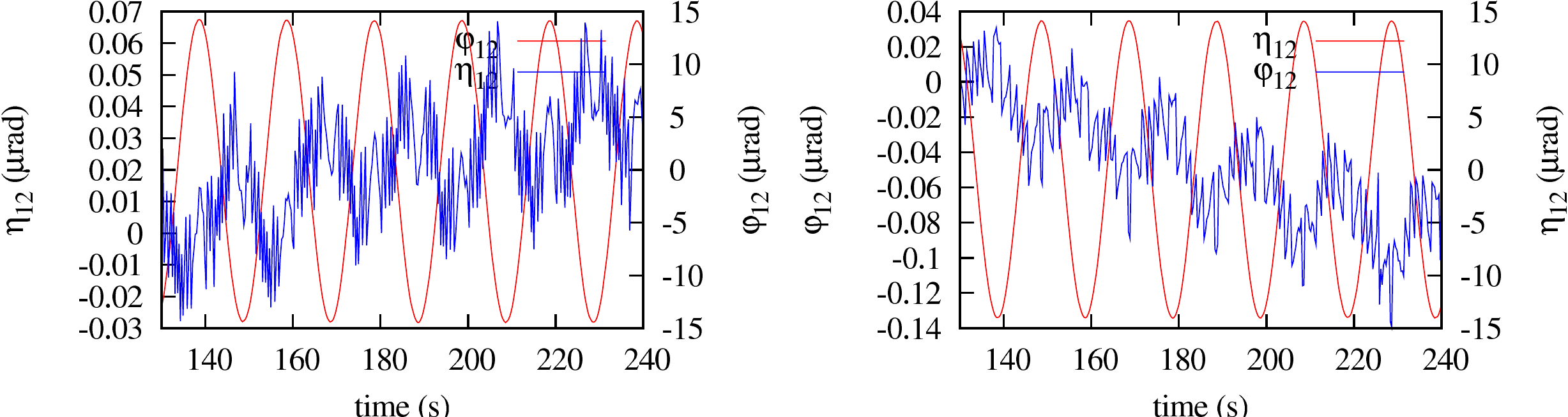}
\caption{Residual coupling of the X12 interferometer angular readout.}
\label{orth}
\end{figure*}

The residual coupling of both test\,mass\,2 angles into each other is shown in Figure~\ref{orth}. In both measurements one angle is sinusoidally modulated by approximately 30\,$\mu$rad and the variations in the second angle are monitored. The coupling for all channels is less than 0.2\,\%, which could still be caused by the piezos used in this experiment. However, it sets an upper limit for the residual cross-talk of the interferometric angular readout.

\subsection{Initial test mass alignment}

\begin{figure*}[htb]
\centering
\includegraphics[width=0.9\textwidth]{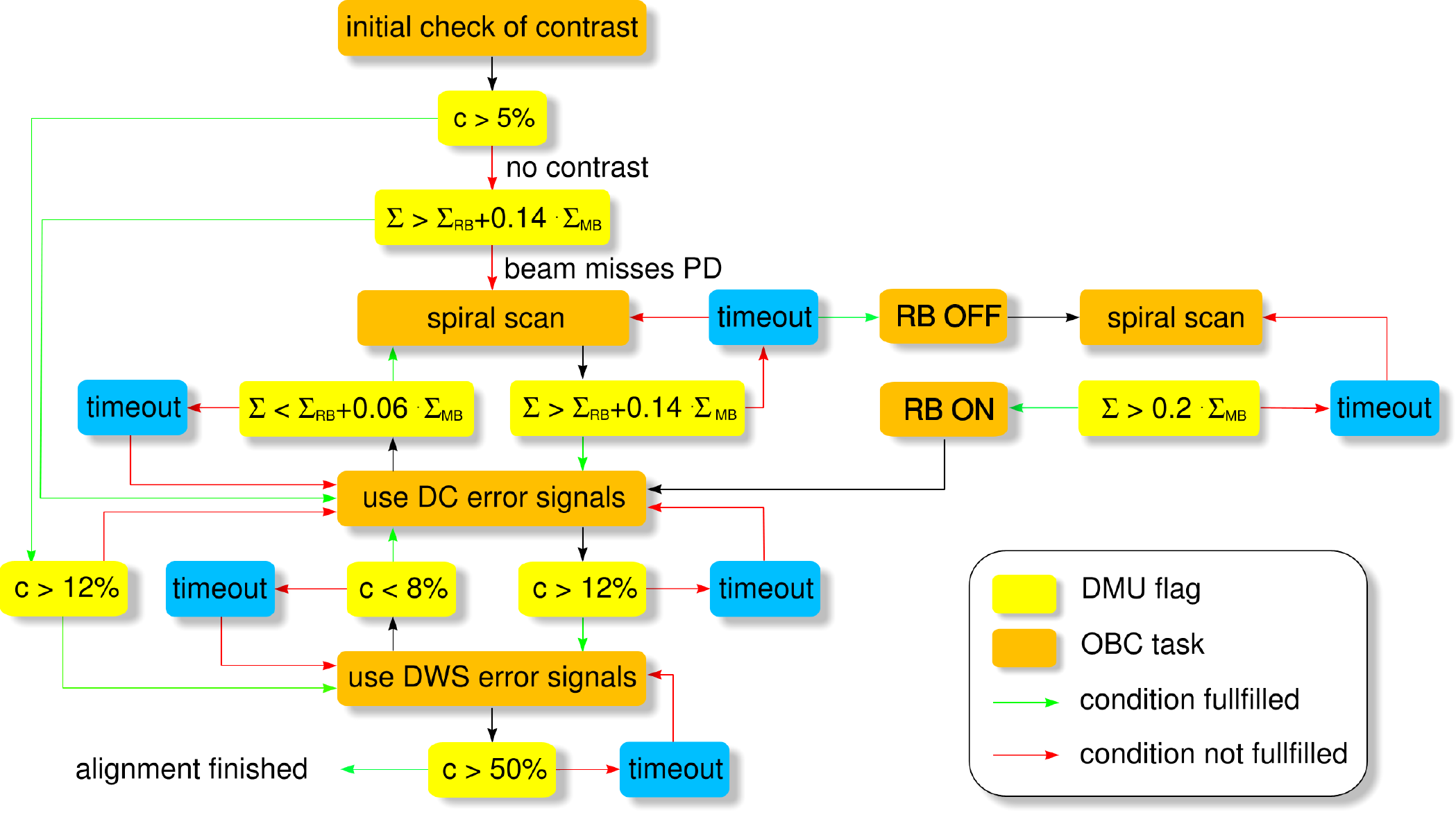}
\caption{Flow-diagram of the test mass alignment procedure.}
\label{alignflow}
\end{figure*}

Within this testbed also the on-orbit initial test mass alignment was simulated, which is illustrated in Figure \ref{alignflow}. A similar procedure has been demonstrated and published in \cite{alignproc}. In the end-to-end testbed some additional modifications were implemented: The processing tasks performed by the DMU and the OBC are separated, wide-range piezo actuated test mass dummy mirrors were used to show the alignment procedure also for large test mass misalignments and the asynchronous data link was implemented as defined in \cite{domenico}.

The alignment procedure must run completely autonomous. The basic idea is to align the test masses to the zero-point of the DWS, where the highest sensitivity of the angular and longitudinal readout is expected. The DWS signals, however, can only be used for the test mass alignment if they are already aligned within a range of about 500\,$\mu$rad which is initially not expected due to the comparatively large static misalignment between the Inertial Sensor Frame and the interferometer frame. Therefore, the test mass is pre-aligned by using the wide-range DC angular readout or, for even larger misalignment, by spirally scanning the test masses.

\begin{figure*}[!htb]
\centering
\subfigure[Angular readout.]{\includegraphics[angle=-90,width=\columnwidth]{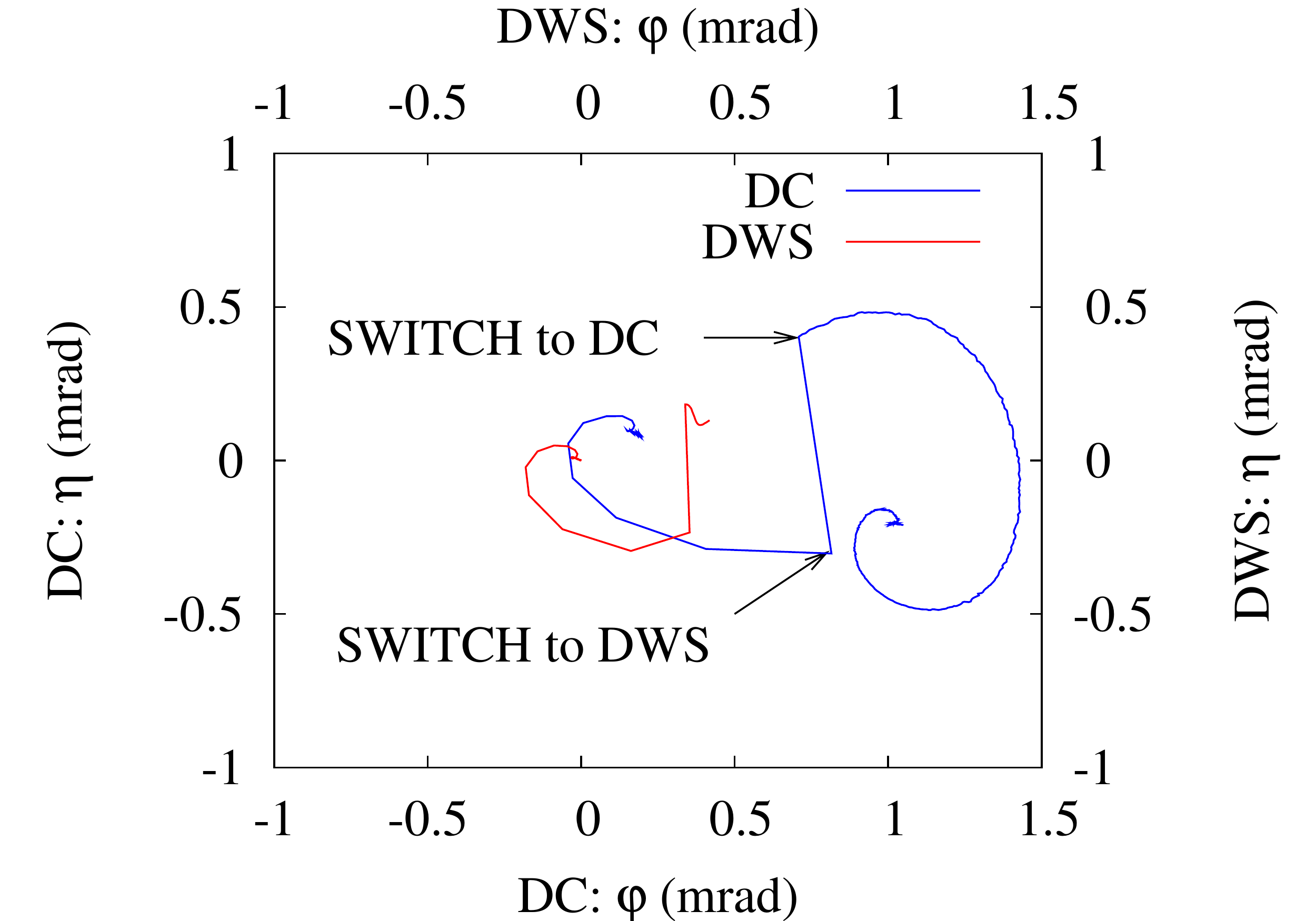}\label{alignres}}
\subfigure[Contrast during alignment.]{\includegraphics[angle=-90,width=\columnwidth]{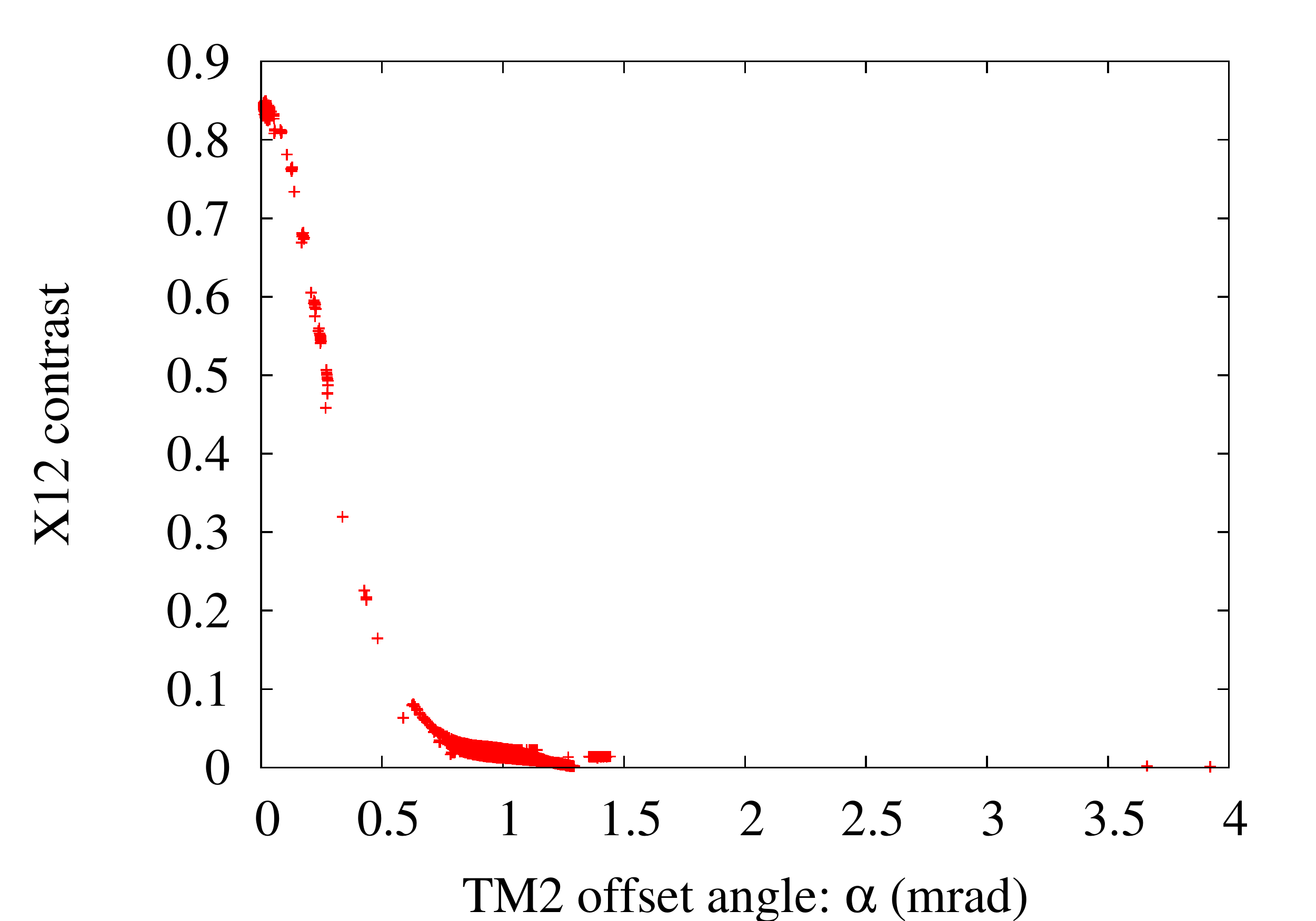}\label{con}}
\caption{Measured angles and contrast during the alignment procedure.}
\end{figure*}

The measured angular readout during the alignment procedure is shown in Figure~\ref{alignres}. It starts with the spiral scan before it switches to DC error signals. When a contrast of 12\,\% is reached, the DWS signals are used for the test mass alignment and the test mass is finally controlled to the DWS zero. For the transition from DC to DWS as error signals for the DFACS, a hysteresis is implemented in order to avoid continuous switching between those signals. Both the DC and the DWS angular readout signals are shown in this graph. Since the centre of the beams do not match with the centre of the photodiode, the DC zero is not the DWS zero.

In Figure \ref{con} the contrast in the interferometer can be seen as function of the beam distance from the DWS zero. It continuously increases with decreasing distance. It can be seen from this figure that the procedure aligns the test mass such that the contrast in the interferometer reaches its maximum.

These results show that the actual design of the test mass alignment procedure is suitable for an autonomous on-orbit alignment.

\section{Acknowledgements}
We gratefully acknowledge support by Deutsches Zentrum f\"ur Luft- und 
Raumfahrt (DLR) (references 50 OQ 0501 and 50 OQ 0601).

\end{document}